Electrical properties of boron-doped MWNTs synthesized by hot-filament chemical vapor deposition

S. Ishii [a,*], M. Nagao [b], T. Watanabe [a], S. Tsuda [a], T. Yamaguchi [a], Y. Takano [a]

[a] National Institute for Materials Science, 1-2-1, Sengen, Tsukuba, Ibaraki 305-0047, Japan

[b] University of Yamanashi, 4-4-37, Takeda, Kofu, Yamanashi 400-8510, Japan

Abstract

We have synthesized a large amount of boron-doped multiwalled carbon nanotubes (MWNTs) by hot-filament chemical vapor deposition. The synthesis was carried out in a flask using a methanol solution of boric acid as a source material. The scanning electron microscopy, transmission electron microscopy, and micro-Raman spectroscopy were performed to evaluate the structural properties of the obtained MWNTs. In order to evaluate the electrical properties, temperature dependence of resistivity was measured in an individual MWNTs with four metal electrodes. The Ramman shifts suggest carrier injection into the boron-doped MWNTs, but the resistivity of the MWNTs was high and increased strongly with decreasing temperature.

Defects induced by the plasma may cause this enhanced resistivity.

PACS codes: 61.48.De, 73.63.Fg, 81.07.De, 81.15.Gh

Keywords: Multiwalled carbon nanotubes; Boron-doping; Hot-filament CVD

*Corresponding author.

Dr. Satoshi Ishii

Postal address: National Institute for Materials Science, 1-2-1, Sengen, Tsukuba, Ibaraki 305-0047, Japan

Phone: +81-29-859-2644

Fax: +81-29-859-2601

E-mail address: ISHII.Satoshi@nims.go.jp

## 1. Introduction

Carbon nanotubes (CNTs) with low resistivity were expected for a variety of applications such as nano wirings in a large scale integration circuit, a SPM probe, a transparent electrode, and so on. Many approaches to control the resistivity using chirarity were attempted because the electrical properties of CNTs strongly depend on the chirarity [1,2]. In our previous work, we have synthesized boron-doped multiwalled carbon nanotubes (MWNTs) by thermal chemical vapor deposition (CVD) using a methanol solution of boric acid as a source material [3-5]. The electrical resistivity of the boron-doepd MWNTs was successfully reduced by increasing the boron concentration up to 2.0 atm% in the source solution. While further doping may reduce the resistivity, growth condition is not yet optimized. In addition to optimizing the synthesis condition, we have to develop the effective synthesis technique to dope the boron further.

In this study, we have synthesized the boron-doped MWNTs by a simple hot-filament CVD technique in a flask. Diamond, which is carbon allotrope, can also be synthesized by the hot-filament CVD [6], though a high-pressure synthesis and a microwave plasma CVD were previously used [7,8]. The plasma induced by the hot-filament with high temperature is expected to decompose the source material containing dopant effectively; it is speculated that more dopant is introduced into the MWNTs.   After the establishment of the hot-filament CVD as the technique to produce a large amount of boron-doped MWNTs with low resistivity, it will accelerate the electrical application of the MWNTs.

## 2. Experiments

Figure 1 shows the synthesizing apparatus that is used in this experiment. The instrument was composed of three flasks connected with each other through stainless steel tubes.   The coiled tungsten filament (1 mm in diameter, 12 cm in total length, 1 mm in loop diameter) was placed in the left flask.   Before the synthesis, the left flask was filled with a source solution.   The middle flask was filled with air, and the right flask contains water for use as a bubbler to prevent the back flow of air.   An Si substrate covered with thermally grown $SiO_2$ (500 nm thick) was placed within 1 mm below the filament.   Before the growth, the surface of the substrate was coated with an Al layer (5 nm thick) by thermal evaporation, and then coated with an iron acetate by dipping followed by heating at 400 $^o$C for 5 min [3-5,9].   When the voltage was applied to the filament, the source solution in the left flask was boiled by the heated filament.   The boiled source solution in the left flask moved into the middle flask through the stainless steel tube pushing the air into the right flask (bubbler), and the air was finally exhausted outdoors.  In this situation, the left flask was filled with the vaporized source solution and the plasma was induced around the heated filament.  As a result, the vaporized source solution decomposed and grew MWNTs on the substrate.  For non-doped MWNTs, the methanol was used as the source material, and the applied voltage and current were 10 V and 7.6 A, respectively.  For the boron-doped MWNTs, the methanol solution of boric acid containing 1.0 atm% of boron was used, and the applied voltage and current were 13 V and 7.4 A, respectively.  For both the MWNTs syntheses, the growth time was 1 hour.

After the growth, the MWNTs were observed using a field emission scanning electron microscope (FE-SEM) at an accelerating voltage of 5.0 kV and a transmission electron microscope (TEM) at an accelerating voltage of 400 kV.  The boron-doping effect on the structural property was evaluated by micro-Raman microscopy with an Ar-ion laser excitation of 514.5 nm in atmosphere and at room temperature.

To evaluate the electrical properties, the temperature dependence of resistivity was measured from room temperature down to 2 K.  For accurate measurement, a four-terminal method was performed with metal electrodes, which were fabricated on an individual target MWNT by electron beam lithography [3].

## 3. Results and discussions

Figure 2 shows the SEM images of (a) non-doped and (b) boron-doped MWNTs synthesized by the hot-filament CVD in the flask. The images show that a large amount of MWNTs could be grown on the surface of substrate. Figure 3 shows the TEM images of (a) non-doped and (b) boron-doped MWNTs. Disorders are observed on the surface of graphen layers of both the non-doped and boron-doped MWNTs. The plasma may damage the structures in the MWNTs synthesized by this technique. However, disorders were reduced by the boron-doping.

Figure 4 shows the Raman spectra taken in the synthesized MWNTs. Two peaks were observed in both the non-doped and the boron-doped MWNTs. The peaks around 1350 cm$^{-1}$ and 1590 cm$^{-1}$ are called as *D* and *G* band, respectively. The *D* band is related to the amorphous carbon or defects and the *G* band is originated from the MWNTs [10]. Therefore, *G*/*D* ratios correspond to the crystal quality of the MWNTs. The *G*/*D* ratios of the non-doped and boron-doped MWNTs were 0.25 and 0.35, respectively; the crystal qualities of both the MWNTs were not good as indicated by the TEM observations (Fig. 3). When the boron was doped, the *G*/*D* ratio increased slightly by 0.1 and the *G* band shifted from 1587 to 1590 cm$^{-1}$. The crystal qualities of the MWNTs were improved by the boron-doping. The upshift of the *G* band indicates that hole carriers were transferred from doped boron to the MWNTs [5,11]. In other word, boron was doped into the MWNTs by the hot-filament CVD technique using the methanol solution of boric acid, although the structural defects were generated.

Figure 5 shows the temperature dependence of normalized resistivity $\rho/\rho_{RT}$; the electrical resistivity was normalized by the value at 300 K. The $\rho/\rho_{RT}$ was increased nonlinearly with decreasing temperature. The $\rho/\rho_{RT}$ of the MWNTs synthesized by the hot-filament CVD was over 5.0 below 50 K, while that of the MWNTs synthesized by the thermal CVD was less than 2.0 even at 2 K as reported in our previous study [5]. These results indicate that the hot-filament CVD technique generated many defects in the MWNTs because the disorders due to the structural defects lead to localization of electrons at low temperatures [12]. The plasma induced around the tungsten filament may generate the defects because the substrate was placed so close to the filament. We have succeeded in synthesizing a large amount of boron-doped MWNTs by the hot-filament CVD, but the defects were generated by plasma and appeared to negate the effect of carrier injection.

## 4. Conclusions

In conclusion, a large amount of boron-doped MWNTs was synthesized very easily by the hot-filament CVD in the flask. The methanol solution of boric acid was used as the source material. The $\rho/\rho_{RT}$ was higher than that of the MWNTs synthesized by the thermal CVD and increased with decreasing temperature [5]. Because the substrate was placed near the filament, the plasma induced by the filament may have generated defects which enhance the resistivity in the grown MWNTs. The effect of the carrier injection was negated by the defects. On the other hand, the defects were reduced by doping 1.0 atm% of boron. The temperature of the substrate was one of the important elements in the MWNTs growth and the filament provided the substrate with the heat. Thus, when the temperature of the substrate is controlled, the substrate can be placed far from the filament, which will prevent the defects and lower the resistivity. We would like to synthesize the boron-doped MWNTs from the source solution containing higher concentration of the boron.

**Figure captions**

Fig. 1 Schematic of apparatus during the synthesis of MWNTs. Three flasks are connected with each other through the stainless steel tubes. The coiled tungsten filament was set up in the right flask filled with the vaporized methanol solution, in which the substrate was placed within 1 mm below the filament. The middle flask works as the buffer to accumulate the methanol solution form the left flask. The right flask works as bubbler.

Fig. 2 SEM images of (a) non-doped and (b) boron-doped MWNTs synthesized by the hot-filament CVD technique.

Fig. 3 TEM images of (a) non-doped and (b) boron-doped MWNTs synthesized by the hot-filament CVD technique. The insets are enlarged images showing the detailed structure of the graphene layers in the MWNTs.

Fig. 4 Raman spectra of the non-doped and the boron-doped MWNTs. The spectra were obtained using 514.5 nm laser excitation.

Fig. 5 Temperature dependence of normalized reistivity $\rho/\rho_{RT}$ in the MWNTs. The solid carves are for the boron-doped MWNTs (B = 1.0 atm%) and the dashed curves are for the non-doped MWNTs (B = 0 atm%).

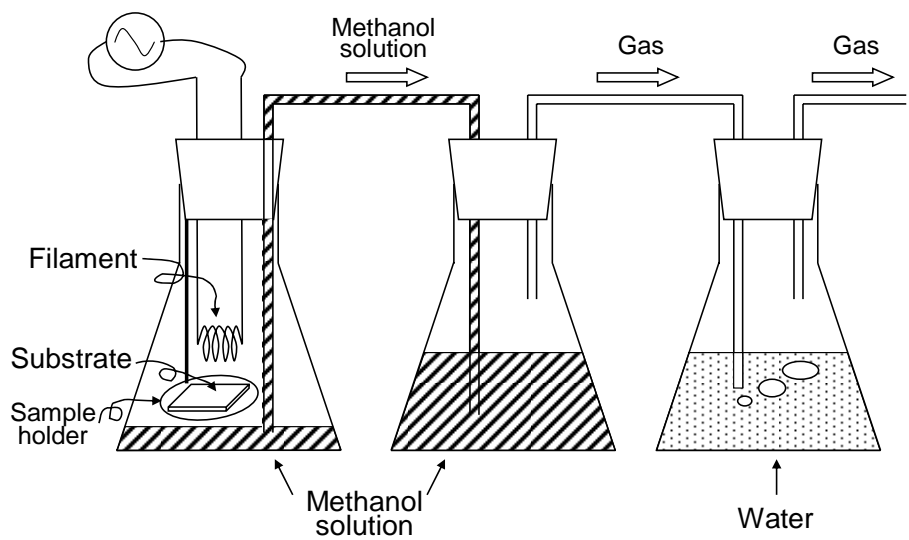

**Figure 1 S. Ishii *et al*.**

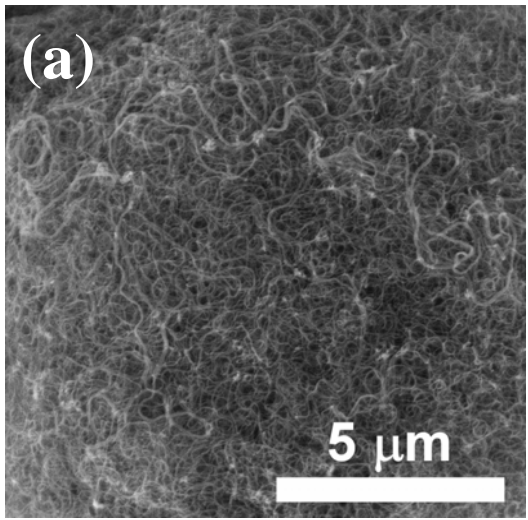 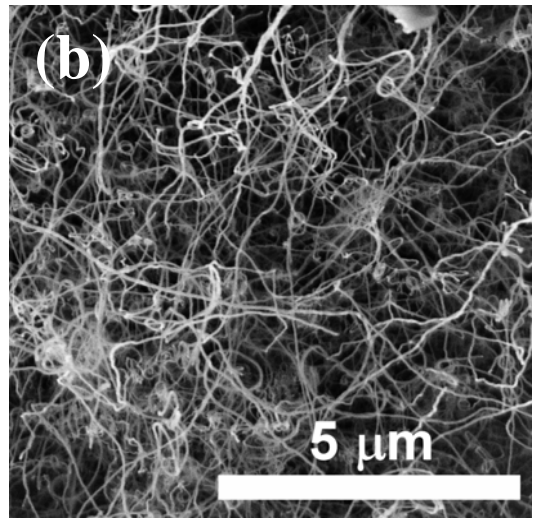

**Figure 2 S. Ishii** *et al*.

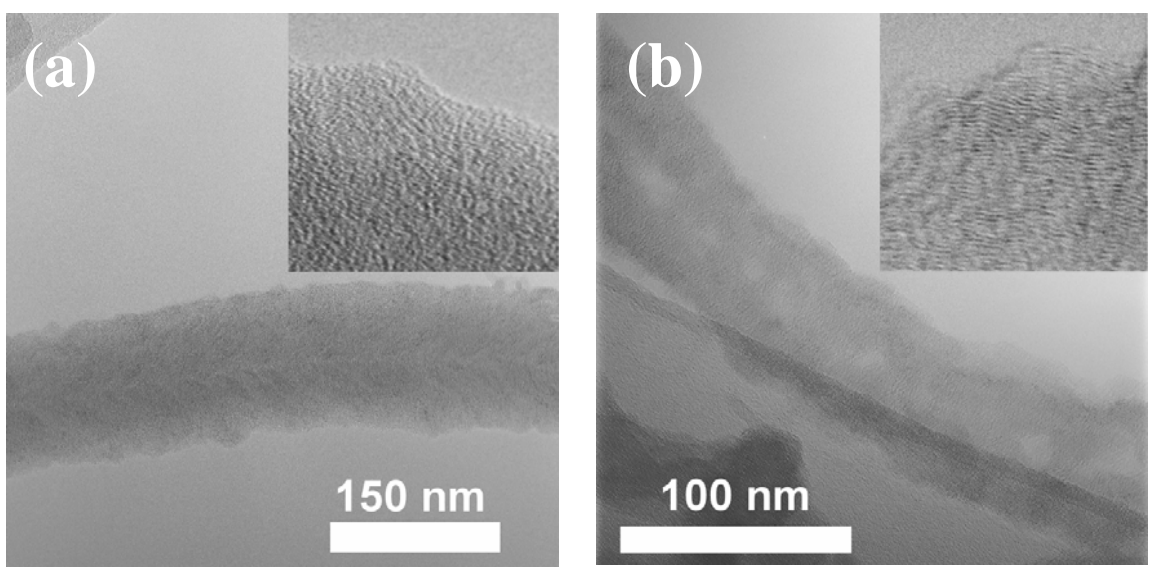

**Figure 3 S. Ishii** *et al*.

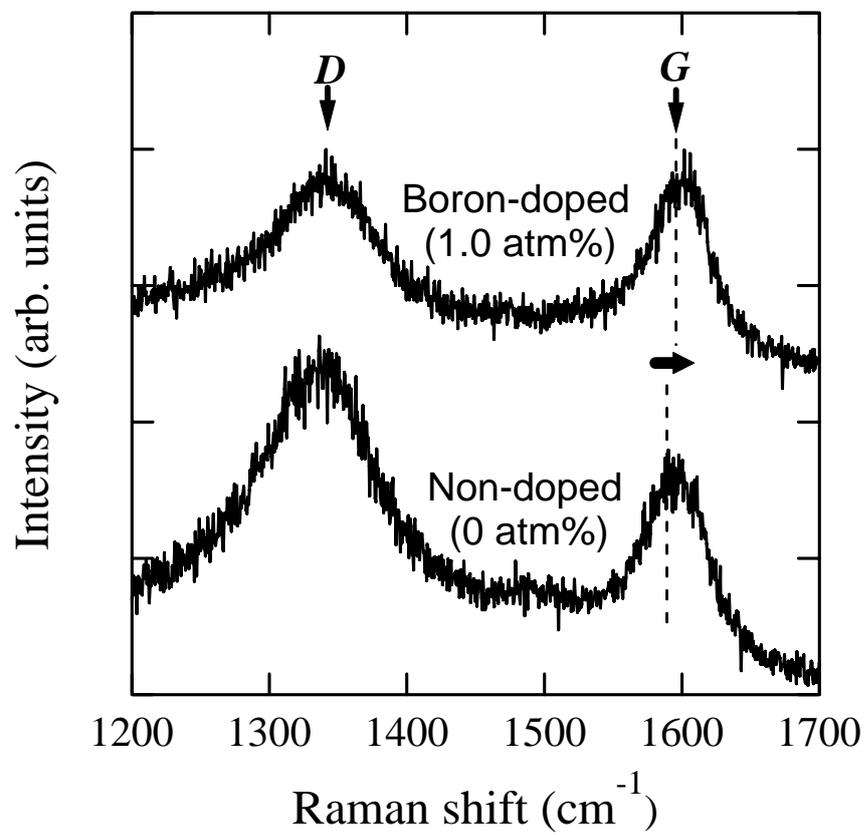

**Figure 4 S. Ishii** *et al*.

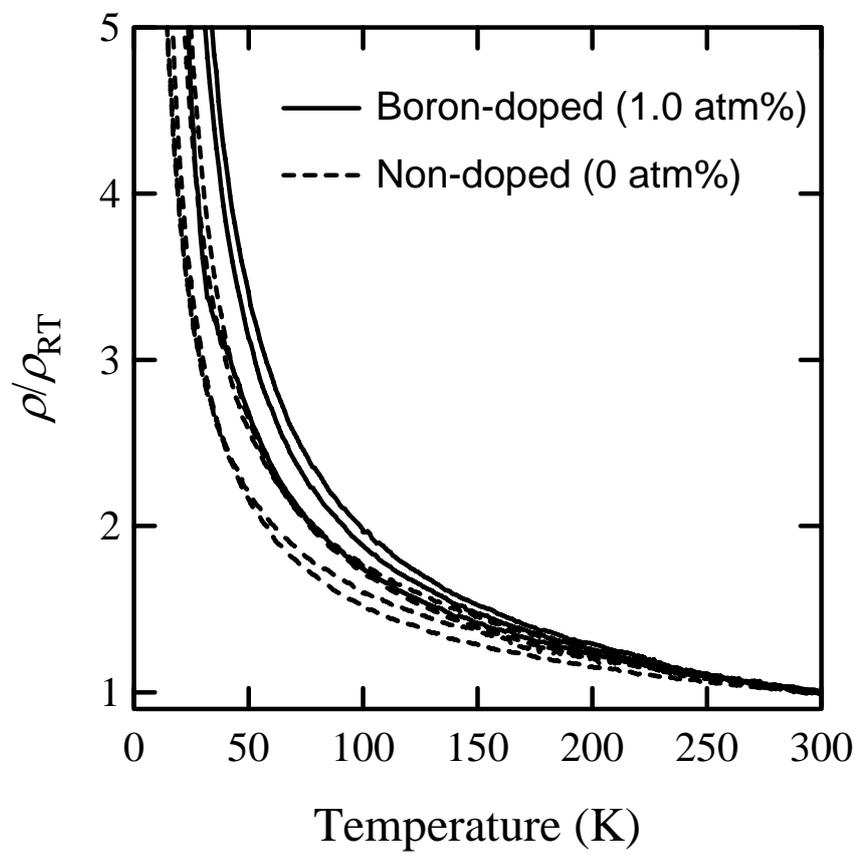

**Figure 5 S. Ishii *et al*.**